# Being, space and time in the Web

Michalis Vafopoulos

**Abstract**

The Web initially emerged as an "antidote" to accumulated scientific knowledge since it enables global representation and communication with minimum costs. Its gigantic scale and interdependence incommode our ability to find relevant information and develop trustworthy contexts. It is time for science to compensate by providing an epistemological "antidote" to Web issues. Philosophy should be in the front line by forming the salient questions and analysis. The scope of our research is to provide a theory about the Web being that will bridge philosophical thinking and engineering. We analyze existence and spatiotemporality in the Web and how it transforms the traditional actualities. The Web space is specified by incoming and outgoing links. The primordial role of visiting durations in Web's existence is approximated by Bergsonian time. The physical space becomes more discoverable. The human activity can be asynchronous, synchronous and continuous. Networked individuals operate in a flexible and spatially dispersed environment. The resulting issues concern the self-determination of a being and the way in which the Web could be a free and open platform for innovation and participation.

WSSC: webscience.org/2010/E.4.1, webscience.org/2010/B.3.2

**Preface**

The fast growing and highly volatile online ecosystem is an integral part of everyday life, transforming human behavior in unexpected ways. Traditional borderlines between public and private, local and global, and physical and digital have started to blur. New types of property, identity and market have been emerged.

Which parts of the existing scientific armory can help us to enrich our understanding as far as the Web is concerned and to build models that anticipate positive future developments?

In order to address such issues we need to overcome common metaphors used in scientific modeling of collective phenomena that can be summarized either in a society encompassing individuals or in an "invisible hand" generating social optimum out of self-contained rational agents or auto-organized collective actions (Latour 2011).

In this paper, two brands of philosophical thinking influence most the proposed metaphors. First, is the Actor-Network Theory (ANT) hypothesis *"not to reduce individuals to self-contained atomic entities but let them deploy the full range of their associates"* (Latour 2011) that the Web makes possible due to the plethora of User datascapes. Second, is the archetypal view of time by Bergson who originated the primordial role of durations. Web User's visiting sessions can be better analyzed as durations, evaluated heterogeneously according to atomic memory and consciousness.

Coincidentally, at their time, both Bergson and Tarde (whose work inspired ANT) had lost scientific debates with regard to the notions of time and society, respectively. However, for both of them, the Web is offering the ground for belated revenge against the arguments of their antagonists.

Bergson during Société Française de Philosophie in 1922 (Bergson 1922) failed to persuade scientific community with his arguments about time as a succession of indeterministic and heterogeneous durations that are irreversible and subject to memory and consciousness (Bergson 1965). Einstein's mathematical proof concerning deterministic physical time prevailed due to the lack of sound scientific riposte. The significance of visiting durations in the Web highlights Bergsonian time as the "time of social systems".

Durkheim, who studied together with Bergson in the École Normale Supérieure, became one of the dominant figures in social research by advocating that society is superior over the sum of its parts and initiated two levels of study between individual psychology and a sui generis societal approach (Latour 2002). Contrastingly, Tarde argued that individuals and society are two sides of the same coin and should not be analyzed separately. Atomic profiles that are massively collected through the Web offer strong empirical evidence for modeling actor and networks as a single entity. Tarde's arguments have started to preponderate over Durkheim's, mainly through the propagation of ANT across scientific borders. It is the Web the place where Bergsonian time meets Tardian approach to individuals and society. The reconstruction of atomic memory and consciousness through the analysis of visiting sessions emphasize the reciprocity of actor and networks in understanding human behavior.

The paper is organized as follows. The first section posses the research questions under investigation and the main results. Being, time and space are analyzed in the section that follows. Original definitions about beings, Web beings, Web time and Web space are also provided. The third section discusses how the emergence of the Web changes the notion of existence, space and time in general. The final section offers a discussion on further implications of the proposed conceptualizations.

1. **Research questions and main findings**

Monnin's call (Monnin 2011) for a more precise description of Web "resources" and Halpin's ontological definitions (Halpin and Presutti 2009a) acted as a good cause so as to investigate the basic constituents of existence in the Web. The present article addresses two research questions:

I.   Which are the main characteristics of being, space and time in the Web if it is considered as a self-contained system that exists in and by itself?
II.  How these idiosyncratic features of the Web transform the traditional conceptions about actual being, space and time?

Web beings are defined as beings that can be communicated through the Web. The URI is the minimal description of invariant elements in this communication and acts like the "fingerprint", the "interlocutor" and the "borderline" of Web beings. It enables direct access, as well as exportation and importation of content. This linking capability, combined with the digital nature of Web being, constitutes the notion of virtualization. Virtualization jointly describes the augmented potentialities of Web being as a digital and distributable unity. The digital facet is discriminated against the physical by five characteristics, namely: nonrivalry, infinite expansibility, discreteness, aspatiality and recombination. The Web, as a distribution network, restricts non-rivalry and infinite expansibility but extends aspatiality, atemporality and recombinance (Vafopoulos 2012).

Location in the Web space is specified by the Web being's URI[1] and the URI's of incoming and outgoing links. These links provide orientation by acting as a three-dimensional "geographic coordinate system" in the Web. The Web space is heterogeneous because Web beings are not characterized by the same network microstructure.

What information consumes is attention. Attention in the Web could be approximated by the visiting time in a Web being. The primordial role of visiting durations in Web's existence qualify Bergsonian time as the best candidate for modeling Web time. The Bergsonian approach of time as successive durations implies indeterminism, heterogeneity and irreversibility. The new property that the Web contributes to the context of "social time" is that it enables the automatic and effortless recording of starting and ending time for all visiting sessions. Durations are becoming discoverable, observable, traceable, processable and massive.

The Web transforms spatiotemporal actuality by adding flexibility and an enriched set of choices for human action. Web space tends to both expand and limit the notion of physical space. Physical space becomes more discoverable and traceable. Transportation and transaction costs become negligible, creating a new range of potentialities. Some of the human activities through or in the Web are available asynchronously, (in part) synchronously and continuously. Networked individuals mobilize part or the whole of their communication system, operating in a flexible, less-bounded and spatially dispersed environment. The resulting issues are related to the self-determination of Web Users and the prerequisites for the Web to be an open platform for innovation and participation.

---

[1] Uniform Resource Identifier.

## 2. Being, space and time in the Web

### 2.1. The facets of the Web

While the Internet has been introduced twenty years earlier, the Web has been its "killer" application with more than two billion Users worldwide accessing some trillion Web pages. Searching, social networking, video broadcasting, photo sharing and micro-blogging have become part of everyday life whilst the majority of software and business applications have migrated to the Web. Web is evolving from a simple fileserver to an enormous database of heterogeneous data. The fundamental hyper-linking property that enables positive network effects in document sharing (Web 1.0) is rapidly expanding to social spheres, contexts (Web 2.0) and URI-based semantic linkages among data (Web 3.0) (Vafopoulos 2011a; Vafopoulos 2011b).

In the present article the term "Web" is used to describe three interconnected parts, namely: Internet infrastructure, Web technologies, online content and Users.

The Web has been initiated as software code of interlinked hypertext documents accessed via the Internet. Using a browser, Users access Web pages that may contain text, images, videos, or other multimedia and navigate among them using hyperlinks.

The Web constitutes an information space in which the items of interest, referred to as resources, are marked up by a set of rules (i.e. HTML), identified by global identifiers (URI) and transferred by protocols (HTTP[2]). They could be considered as *"contemporary forms of what the Greeks of antiquity called hypomnemata"* (i.e. mnemotechnics) (Stiegler 2010). The Web has become the most successful and popular piece of software in history because it is build on a technical architecture, which is simple, free or inexpensive, networked, based on open standards, extensible, tolerant to errors, universal (regardless of the hardware and software platform, application software, network access, public, group, or personal scope, language, culture operating system and ability), powerful and enjoyable.

Human participation in massive scale upgraded the Web from a piece of software to a "social machine" (Hendler et al. 2008). The Web has been evolved to a highly interdependent living ecosystem, which affects Users and non-Users in everyday and some life-critical choices. Last decade, the transformative power of the Web stimulated its study as a standalone techno-social ecosystem under the umbrella of Web science (Berners-Lee et al. 2006; Vafopoulos 2011c).

---
[2] HyperText Transfer Protocol.

### 2.2. Being in the Web

The notion of "resource" has been selected to describe the underpinning unit of the Web ecosystem. Rarely, it is interchanged by terms like "object" and "thing" or others. Basically, the concept of "resource" carries strong economic and ecological connotations. In Economics, the basic resources are three: land, labor, and capital. The term "resource" enjoys twelve appearances in the scientific classification of Economics[3]. In broader sense, they are human, natural and renewable resources. Additionally, in the relevant Wikipedia entry there is not a single word about Internet or Web[4]. On the other hand, terms like "object", "entity" or "thing" are too general and not descriptive enough since they convey multiple meanings under different contexts. Thus, a new concept for the Web's cells is required to demonstrate the coevolution between technological and social aspects; its living nature consisted of internal laws of existence, dynamism and transformational impetus. It is important to introduce existence in the Web, based on a clear, minimal and pragmatic definition, which describes the Web as an integral part of the world and could be useful for Web scholars and engineers from diverse origins.

This effort is initiated by our baseline definitions about being in general and Web being in particular.

***Being:*** *A being exists if and only if there is a communication channel linking to it.*

This definition is not only practical but also general enough to include many existing theories in philosophy. For instance, it could be considered an alternative and minimal approach to "Dasein", as it was re-introduced by Heidegger (Heidegger 1962). Latour explains, *"Dasein*

---

[3] http://www.aeaweb.org/jel/jel_class_system.php
[4] http://en.wikipedia.org/wiki/Resource

*has no clothes, no habitat, no biology, no hormones, no atmosphere around it, no medication, no viable transportation system... Dasein is thrown into the world but is so naked that it doesn't stand much chance of survival"* (Latour 2009).

The concept of "communication channel" in defining existence, encapsulates the core idea of ANT about "circulating entities" and the useless dichotomy between individuals and society that is commonly the case across various disciplines (Latour 2005). In the Web era, "to have" (e.g. friends, connections, identities) becomes of equal importance and should be studied along with various manifestations of "to be". One aspect of being in the Web is that the communication channel is more concrete, identifiable and visible. URI is the fundamental technology on creating communication channels in the Web.

**URIs**[5] (e.g. http://www.vafopoulos.org) are short strings that identify resources in the Web: documents, images, downloadable files, services, electronic mailboxes, and other resources. They make resources available under a variety of naming schemes and access methods such as HTTP and Internet mail addressable in the same simple way. They reduce the tedium of *"log in to this server, then issue this magic command..."* down to a single click. Every URI is owned by a physical or a legal entity, which has the right to sell it or to provide access to any other entity she wishes (refer to (Halpin and Presutti 2009b) for a thorough discuss in identity and reference on the Web.

**Web beings**: *Web beings are defined to be beings that can be communicated through the Web.*

URI is the minimal description of invariant elements in communication through the Web and characterizes unambiguously the Web being. It is like the "fingerprint" of the Web being because is directly connected to existence (birth, access, navigate, edit and death). All the other characteristics of Web beings may alter in time (e.g. appearance, content, structure), but a change in URI means the death of an existing and the birth of a new Web being. The main role of URI is to categorically identify the Web being and to enable navigation and recombinance over the network of Web beings.

---
[5] http://www.w3.org/Addressing/

Web beings can be information-based or non information-based (as resources defined by (Halpin and Presutti 2009a)), but in both cases are named, referred and identified through, at least one exclusive URI, which constitutes the minimum amount of digital information.

The question is how a being can be artifactualized (Monnin 2009) to a Web being. In particular, research is concentrated on how digitality is mutated by the linking potential, enabling beings to be anywhere, at anytime.

### 2.3. Virtualization = digital + linking

The Web revolution is based on the realization of publishing, inter-connecting and updating digital objects with low cost and effort globally. The massive re-allocation of action over distributed datascapes takes place because digital is augmented and ramified by linking capacities. Intuitively, every User can access all available Web beings at anytime from anywhere. We are all "potential" (or "quasi") owners of each Web being, in the sense that it may not reside in our memory device but can be downloaded almost instantly. This fundamental expansion of existence can be better captured by the concept of *virtualization*. Virtualization jointly describes the augmented potentialities of a Web being as a digital and distributable unity. Linking is based on digital but is metamorphosing it into a more volatile, open and multi-functional artifact. According to Lévy (Lévy 1998): *"Virtualization is not derealization (the transformation of a reality into a collection of possibles) but a change of identity, a displacement of the center of ontological gravity of the object considered…. The real resembles the possible. The actual, however, in no way resembles the virtual. It responds to it. …Rigorously defined, the virtual has few affinities with the false, the illusionary, the imaginary. The virtual is not at all the opposite of the real. It is, on the contrary, a powerful and productive mode of Being, a mode that gives free rein to creative processes."*

### 2.3.1. The digital aspect

The digital aspect of existence has been extensively investigated by diverse research disciplines (see for instance (Negroponte 1995; Negroponte 2000; Kim 2001; Castells 2003)). Digital beings are beings that are constructed by sequences of bits, 0s and 1s. They are usually created by mental efforts and occupy negligible parts of physical space. By design, they cannot directly satisfy basic needs like food, water, shelter and clothing. The most relevant analysis emerges if we focus on "digital goods", which are digital beings with economic value. Quah (Quah 2003) distinguishes digital from physical goods by five characteristics, namely: nonrivalry, infinite expansibility, discreteness, aspatiality (or weightlessness or spacelessness), and recombination. Particularly, in economic terms, digital goods are:

(1) *Nonrival.* Most physical goods are rival, in the sense that consumption by one consumer diminishes the usefulness to any other consumer (e.g. if I drink a bottle of water, I exclude everybody of drinking this particular water). Contrastingly to most physical goods, digital goods are nonrival, in the sense that many Users can consume videos and philosophical theories without preventing their use from others. The concept of nonrivalry plays an important role in economic analysis ((Rivera-Batiz and Romer 1990; Shapiro and Varian 1999)).

(2) *Infinitely expansible.* According to Pollock (Pollock 2005): *"A good is infinitely expansible if possession of 1 unit of the good is equivalent to possession of arbitrarily many units of the good - i.e. one unit may be expanded infinitely. Note that this implies that the good may be "expanded" both infinitely in extent and infinitely quickly".* Only in the limit case of perfect nonrivalry is equivalent to infinite expansibility. Infinite expansibility and virtually zero-cost copying and distribution of digital beings, are strong changing forces in the business model of media industries.

(3*) (Initially) discrete or indivisible.* Digital goods are (initially) discrete, in the sense that their quantity is measured exclusively by integers, and are only instantiated to quantity one. Alternatively, digital goods are not divisible. As Quah (Quah 2003) explains: *"Making a*

*fractional copy rather than a whole one, where the fraction is distant from 1, will destroy that particular instance of the digital good."* Indivisibility can be further refined by the properties of fragility and robustness. In economic terms, a digital good is robust if a sufficiently small and random reassignment or removal of its bit strings is not affecting the economic value of the good (e.g. MP3 compression). Elseways, the digital good is defined to be fragile (e.g. computer software).

(4) *Aspatial*. In the limit, digital goods are aspatial in the sense that they are everywhere and nowhere at once. Aspatiality of digital goods is not identical to the definition of aspatiality in Plato's theory of Forms or theory of Ideas (Ross 1953), which refers to the absence of spatial dimensions, and thus connotes no orientation in space. Digital goods are real bits located in physical devices (e.g. hard disk).

(5) *Recombinant*. Despite the majority of ordinary goods, digital goods are *"recombinant, cumulative and emergent-new digital goods that arise from merging antecedents have features absent from the original, parent digital goods."* (Quah 2003).

### 2.3.2. The linking aspect

The above-described characteristics of digital goods can be proliferated through a compatible distribution channel, which enables complementarity in creation, exchange and access. The most efficient channel has proven to be the Web network. In a few years time, the Web has evolved into an enormous repository and distribution channel of information. Web technologies provide the technical platform for representing, identifying, interconnecting and bartering addressable information. In the Web, the digital becomes tangible, editable, uniquely definable and compatible to almost any format.

A cornerstone in this process is URI technology. URI is the specific part of digital information contained in a Web being that identifies uniquely and enables direct linking and transfer to other Web beings. Every single URI identifies exclusively one Web being. Although each Web being has one generic URI, it can be identified through many other URIs. URI is the "fingerprint", the "borderline" and the "interlocutor" of Web beings. It facilitates the "teleportation" of navigators (i.e. direct access), as well as the automatic exportation (e.g. RSS feed) and importation of content from other Web beings (e.g. inclusion of twitter streams in personal page). The latter emerges, in the most concrete way, as a trade off between original creation and reference to existing work. An old dilemma, mainly existing in scientific research, returns as a practical question in creating and commercializing Web beings.

The Web, as a distribution channel, restricts non-rivalry and infinite expansibility of digital goods because of its limited concurrency capacity and costs imposed by the underlying infrastructure. Digital goods in the Web are initially discrete and indivisible like their underlying digital goods. Their distinctive characteristic is the facilitation of massive recombinance and consumption in micro-chunks. At the current Web 2.0 era, Users can easily edit, interconnect, aggregate and comment text, images and video in the Web. Most of these opportunities are engineered in a distributed and self-powered level. Conjointly, to massive information aggregation and recombinance, the Web extends aspatiality and atemporality of digital goods. Distributed digital information is characterized, not only by the fact that enjoys low transportation cost, but also by its accessibility from anywhere, anytime. Actually, the Web expands aspatiality and atemporality from local level (e.g. personal hard disk) to global level (e.g. downloadable file link).

### 2.4. The Web Space is the online network

Every Web being could be accessed directly by typing a URI in browser's address bar or indirectly, through search engines or clicking on referral Web beings. Analogously, to the physical world, the Web space could be considered as a division of position and place of online content, created by the links among Web beings. Each Web being occupies a specific locus in the Web network. The identification in the Web space is given by the URI namespace, whereas the location is specified by a triplet of URIs: the Web being's $URI_i$ and two sets of URIs: incoming ($URI_{ji}$) and outgoing links ($URI_{ij}$) of the Web being. These links provide orientation by acting as a three-dimensional "geographic coordinate system" in the Web.

$$Web\ Space\ =\ \mathcal{S}\ (URI_i, URI_{ij}, URI_{ji})$$

The Web space is heterogeneous because Web beings are not characterized by the same network microstructure. For instance, vafopoulos.org and w3c.org have different sets of incoming and outgoing links, and thus, different pathways and possibilities to be accessed. The study of network structure in the micro and macro level is an important issue in many disciplines like mathematics and physics, computer science, network economics, game theory and social studies (refer for example to (Newman 2008), (Jackson 2008) (Albert and Barabási 2002), (Watts 2003)). The calculation of network metrics (e.g. hubs and authorities, betweeness and eigenvector centrality) and algorithms (e.g. community detection) model under different criteria the "gravity" and the relative "distance" among beings in the Web space. Likewise, eigenvector centrality was the initial base of Google's search algorithm (Page et al. 1998) that successfully assists us to explore the Web space and to locate Web beings.

How does the Web space evolve over time? The creation or the deletion of a single Web being or link constitutes a change in Web space. Consequently, the evolution of Web space is fully described by the birth and death time processes of Web beings and URIs.

## 2.5. The Web Time is Bergsonian

### 2.5.1. Attention: the "currency" of the Web

Navigation in Web space results in traffic. Web traffic is defined, as the successive activation of URIs that is recorded in Web being's log file. In order to prevent confusion, has been established the convention that events in the Web are synchronized under the UTC[6]. Actually, this is the first time that humanity has established a universal event log system to such extent, scalable and heterogeneous socio-technical system. The Web's embedded logging system records minimally the access date and time, User's IP address, the viewed URI, the amount of transferred bytes and verification about the status of the transfer[7]. For analytical purposes, the initial logs are aggregated with respect to Users or Web beings. In the first case, User sessions are formatted as temporally compact sequences of visits in a series of Web beings by a specific User. Alternatively, User sessions can be aggregated with respect to Web beings (for related issues in modeling Web traffic see (Abdulla 1998)).

---

[6] Coordinated Universal Time.

According to the original Web architecture, the owner of a Web being has the exclusive access in the log file, which describes the traffic of Users in the local Web space. Lately, many companies extend their control in Web space by collecting traffic data outside their Web beings (e.g. cookies, surveys). In most cases, this expansion takes place without the consensus of Users, raising important issues and debates about privacy and self-determination in the Web (Goldfarb and Tucker 2010), (Angwin 2010).

In order to understand the central role of visiting durations in Web's function, let us investigate the inner information flows, assuming that the Web is a self-contained system that exists in and by itself. Web Users are navigating or/and producing Web beings. Thus, Users are partitioned to Navigators and Editors of Web beings. Navigators explore the Web to acquire utility by consuming Web beings[8] (Figure 1). This navigation creates traffic streams for the creators of Web beings (Editors). The main motivation for creating and updating Web beings is to attract visitors. Non-professional Editors provide information, effort and time for free, mainly because of individual acclaim and reputation-building, moral reward and self-confidence. In some cases, this temporal disengagement between effort and reward is explained by an expected increase of Editors' choices to high-paying employment arrangements. Contrastingly, professional Editors of Web beings are profit maximizers and take into account direct financial compensations (e.g. a blog with paid advertisements). In both cases, the time that visitors spend in Web beings (sessions) is crucial for their existence because is directly related to Editors' expectations. The resulting income acts as an incentive for professional Editors to update the already existing and create new Web beings, producing a new Web space with novel possibilities for Navigators in order to maximize their utility.

---

[7] http://www.w3.org/Daemon/User/Config/Logging.html#common-logfile-format
[8] In Web economy analysis, Web beings can be specifically defined to be "Web goods". "Web goods" are Web beings that affect the utility of or the payoff to some individual in the economy 5/4/12 6:47 AM.

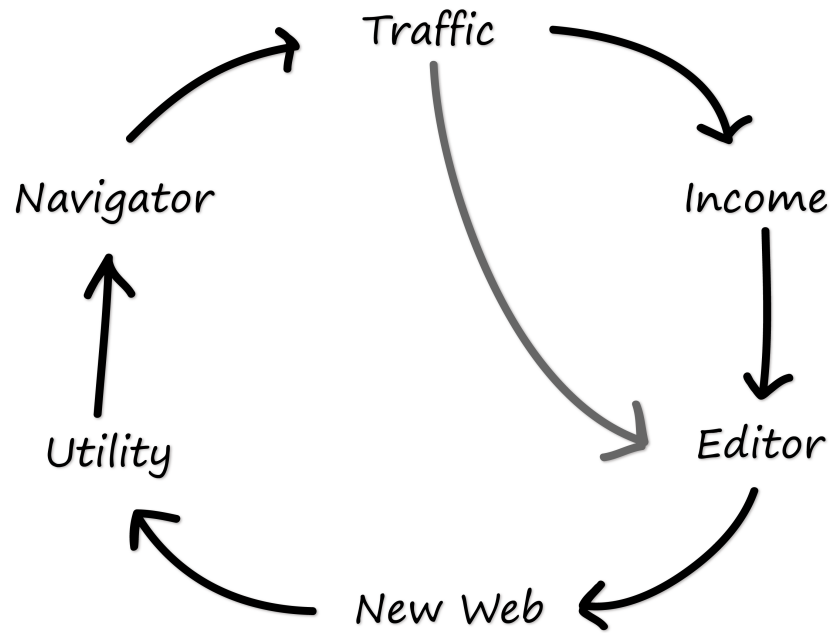

Figure 1: The inner function of Web economy: Navigators explore the Web to acquire utility, resulting exploitable traffic for Editors, which are motivated to update the existing Web.

What information consumes is attention (Simon 1971). Attention in the Web could be better captured by the visiting time in a Web being. The monetization of visiting sessions has created the biggest internal Web market with more than 63 billion dollars turnover in 2010 (Yarow 2011). Hence, attention could be considered as the "currency" of the Web and visiting durations as the exchanging unit in the Web market.

### 2.5.2. Visiting sessions: the time of the Web

Apart from the "book-keeping" clock time that is strictly defined in physics, time could be described as a series of choices in space. Analogously, Web time is a series of choices (visits) in the Web space that can be better modeled as Bergsonian durations, because visiting selections attach semantic meaning and define casual relationships among Web beings.

The Aristotelian view asserts that time is like a universal order within all changes are related to each other. Time is meaningless if there are no tangible events (Aristotle IV 218b 2007). Going many steps further, Bergson defined time as a sequence of finite and heterogeneous durations. Time is irreversible and the future unpredictable (Bergson 1910). Each duration has different significance from the preceding and the following ones. The transition from inner time to the time of things is related to *memory* and *consciousness*. As Bergson explains, *"Without a ultimate memory, which links two instants, the one to the other, there would be only the one or the other of the two, consequently a unique instant, without before and after, without succession, thus no time. …It is impossible to distinguish between the duration, however short it may be, which separates two instants and a memory that connects them, because duration is essentially a continuation of what no longer exists into what does exist. This is real time, perceived and lived. This is also any conceived time, because one cannot conceive a time without imagining it as perceived and lived. Thus, duration implies consciousness; and we place consciousness deeply in the things by crediting them a time that endures"* (Bergson 1965).

In the same line, Aristotle wonders *"…would there be time if there were no soul?"* (Aristotle IV 223a 2007) and remarks *"When we are conscious of no mental change we do not think time has elapsed, any more than the fabled sleepers of Sardinia do when they awake; they connect the later now with the earlier and make them one"* (Aristotle IV 218b 2007).

Contrastingly to the deterministic clock time of integrable systems, the Bergsonian time describes the temporal processes of irreversible dynamic systems. Deterministic models are unable to explain the clock time irreversibility and thus, to anticipate the temporal complexity and dynamism of real systems. The Bergsonian approach of time as duration is characterized by indeterminism, heterogeneity and irreversibility, capturing the essence of human behavior. We argue that the Bergsonian time is the "time of social systems", and is built on the Einsteinian time of physical systems.

The new property that the Web contributes to the context of "social time" is that it enables the automatic and effortless recording of starting and ending time for all visiting sessions. Durations are becoming not only discoverable and observable but also traceable, processable and massive. These online visiting durations increase the material dimension of networks.

Bergson's belief on time irreversibility stems from the general irreversibility of a being. In the Web, the being is related to time in the sense that what it means for a Web being to be, is to exist temporally in the stretch between birth and death. While birth of a Web being in physical time is the action of uploading and assigning an exclusive URI, birth in Web time comes from the first visitors. Analogously, death in physical time is the removal of Web being's URI and in Web time the full absence of visitors.

Recently, Antoniou and Christidis (Antoniou and Christidis 2010) have made the first step in formalizing Bergsonian time by introducing the relevant time operator in terms of innovation and aging. Their approach is initiated by the definition of clock time as a scalar quantity, the values of which correspond to a subset of real numbers. The induced "age" of a random variable is the "age" of each time duration weighted by the probability of innovation, though they have not incorporated the significance of durations with consciousness and memory.

In general, the sustainability of existence in the Web is described as a cyclical economic function on the inner Web space and time (Figure 1 in subsection 2.5.1). In this function, some business and governments (e.g. in China and Turkey) try to take control of Web space and time. Their usual approach is to model human behavior by reconstructing Users' memory and consciousness based on their visiting sessions. For the moment, a small number of mammoth companies control most of the Web space and time, restricting governments to passive and repressive policies (Vafopoulos 2012). Google, Facebook, Amazon and Apple, to name only a few, based on vast amounts of collected and processed User data, act as gatekeepers of the online ecosystem. This fact undercuts the foundations of the Web and has started to raise severe remonstrations from the academic community. The "computational social science argument" (Lazer et al. 2009) can be summarized to the substantial barriers which are posed to scientists, mainly from business, to collect and analyze the terabytes of available data, describing minute-by-minute interactions and locations of entire populations of individuals.

### 3. How the Web affects traditional Space, Time and Being

#### 3.1. Space with the Web

Let us now relax the hypothesis of the self-contained Web and describe how the Web affects physical space, time and existence.

User logs of every visiting session register User's Internet Protocol address that can be used to trace back the Internet Service Provider and computer terminal. By this way, the Web space is directly and automatically connected to the physical space.

The Web space assists us to navigate the physical space and vice versa. In that way the physical space becomes more *discoverable* and *traceable*. Many Users find their destination through online maps and farmers make decisions for their plants based on online weather forecasts. Google is now adjusting search results to the declared place of origin by the Users for related queries. Generally, Web space tends to both expand and limit the notion of physical space. Hyper-connected Users may carry their digital "Little Box" anywhere but it can also be the case that some of them may restrict their transportations due to the role of the Web as communication replacement (Wellman 2002). Apart from massive navigation, aggregation and recombinance, the Web extends *aspatiality* of existence. In economy, geography still matters and especially in the production of knowledge-intensive industries, whose synchronous face-to-face interactions and critical mass in human capital are important inputs. The major implication of aspatiality is that transportation and transaction costs are negligible, creating a new range of possibilities.

Stigler explains that *"Post-globalization is not a territorial withdrawal: it is on the contrary the inscription of territory in a planetary reticularity through which it can be augmented with its partners at all the levels of which it is composed, from the interpersonal relation made possible by the opening up of rural regions implementing a politics of the digital age, to business which, deploying its competence locally and contributively, knows how to build a de-*

*territorialized relational space: ecological relational space is a territory of hyper-learning…"* (Stiegler 2010).

### 3.2. Time with Web

Before the advent of the Web, the notion of "market" was mainly used to describe physical meetings between sellers and buyers in order to exchange goods for a price, in a specific time frame. Traditional marketplaces were based on synchronization and agreement over time, place and price.

In the Web market, participants can make decisions across a wider spectrum of choices. They cannot smell or touch but they can search for quality and price in 24/365 basis. Most of human activities that take place through or in the Web have become available *asynchronously*, (*in part*) *synchronously* and *continuously* (e.g. e-commerce, social networking).

If physical time is an arbitrary standard that enables the division of infinite space to useful parts, then the Web assist us to separate it in more fine pieces. What the Web contributes in the physical time and space is *flexibility* and an enriched set of choices for human action. An increasing number of individuals, groups and businesses reclaim this flexibility, putting pressure in traditional socio-economic structures, which are based on the pre-Web physical time and place hypotheses.

### 3.3. Being with the Web

This resulting paradigm shift has circular effects on personal choice and collective action. In particular, proliferation of wireless Internet and mobile devices have contributed to the transformation of human communities from the densely knit "Little Boxes" (linking door-to-door) to the sparsely knit "glocalized" networks (linking both locally and globally) and to "networked individuals" (linking with little regard to space) (Wellman 2002).

Networked individuals mobilize part or the whole of their information and communication system, operating in a more *flexible*, *less-bounded* and *spatially dispersed* environment. This innovative modus operandi includes frequent *switch* among multiple social networks and modes of communication, different sense of belonging, flexible business arrangements and intense time management.

Emergence of the Web ecosystem increased the quantity and quality of opportunities not only to discover and update information but also to communicate. The quantity could be simply approximated by the largest amount of links among beings and Web beings. This impressive augmentation of physical space with Web space capacities has also affected the quality of human interaction. The Web broadens face-to-face communication by expanding the reach of social networks to multiple-channel communication. As Wellman (Wellman 2002) explains the Web *"...simultaneously affords: (a) personal communications between one or multiple friends, (b) within network broadcasts; and (c) public addresses to strangers."*

The catalytic role of the Web in providing new opportunities for collaboration and creativity has been crystalized under the development of a new third realm, the *privatised space*. The privatised space arises between the private realm of intimacy and individualism and the public realm of citizenship and active participation for the societal good (including professional activity). The Web functions mainly as a privatized space with public life, sociability and public opinion, with public interactions and visibility, but private reasoning and motivation (O'Hara and Shadbolt 2008).

In this new privatazed space, has emerged a decentralized Peer production through loosely affiliated self-powered entities as the third mode of production, a third mode of governance, and a third mode of property (Bauwens 2006). Peer production communities are based on information sharing mechanisms so as to create public knowledge repositories and encapsulate community's aggregated preferences and expectations. It constitutes a new form of decentralized inter-creativity outside the traditional market and price mechanisms by redefining two economic orthodoxies: *diminishing marginal productivity* and *increasing returns to scale* (Vafopoulos 2012). Peer Production is considered to be a new form of production that complements the existing and extends David's taxonomy adding the fourth P (Property, Procurement, Patronage and Peer Production) (David 1992).

As the distributed action enhances and partially substitutes groupware, the issues of security, privacy, identity and trust in massive scale become of primary importance.

### 3.4. Discussion

The Web transforms the way we work, shop, learn, communicate, search, participate, and recreate, introducing unexpectedly novel and complex actualities. Immaterial thoughts and revealed preferences are becoming visible, describable and traceable parts of collective existence. However, is yet difficult to fully understand and anticipate the cumulative effects of these extra choices to human societies. Engineering, business and regulatory decisions about the Web often result in unanticipated outcomes. The questions about the viability, feasibility and efficiency of the Web engage higher spots in many research agendas. Pessimists and optimists, technophobians and technophiles, engineers and social scientists, bloggers and CEOs question the present and the future of the Web with equal zeal.

Really, what changes are required so that the Web can work better?

The contribution of philosophical thinking in this inter-disciplinary dialogue could be threefold: (a) to establish a common minimum understanding about existence, time, space and the underlying values of the Web, (b) to identify the important issues arising from the Web and (c) to build models and metrics that will enable decision making about the Web. This campaign will be successful if Web scholars from diverse disciplines incorporate in their research projects, thoughts, arguments and mechanisms related to the above questions.

The present paper addresses (a) by defining being, space and time in the Web. The Web "curves" physical time and space by adding flexibility, universality, more available options and sources of risks. Its growing influence proliferates the available data to construct detailed User profiles and related theories of behavior and collective action.

The emerging function of virtualization creates new possibilities and redistributes human choice, resulting in more complex relations. The plethora of (almost) continuous User data, enables the joint analysis of entities and their attributes. In this context, Web scholars and regulators face two major strands of challenges:

- to obtain the right balance between open access to online information and self-determination of Users, on the one hand, and provide the proper incentives to produce content and develop network infrastructure on the other, and
- to accelerate socio-economic development by facilitating life-critical functions in the developing world and by enabling transparency, participation and added-value services in the developed world.

The aforementioned issues entail a series of second-level components that further elaborate their diverse facets. These components may refer to matters like the lack of credibility and quality of some Web beings, propagandas, defamation, rumors, addictogenic behaviors (Stiegler 2010), waste of time of Users and so forth.

The next steps in addressing this new form of complexity could be directed in describing the core functions of beings and Web beings under a computable framework, which builds on results in network and Web analysis.

We believe that the Web ecosystem is an ethically-relevant social machine and it should be systematically analyzed as such, in order to realize its potential in promoting human values. In fact, the initial motivation behind the development of the Web itself was esteem, pride, excellence, absence of guilt, rewards, and indignation (O'Hara 2010). The ethical analysis of the Web as an integral part of philosophical studies, should investigate how basic values like freedom are transformed through and in the Web.

As in physical space, the ultimate value in the Web could be the ensurance of free will of Users to collectively transform Web beings (Wark 2004). The ongoing change in paradigm alters the content and means of achieving self-determination in and off the Web. Being self-determined in the Web is strongly related to underlying personal data management processes. These processes should assure anonymity and lethe in particular contexts, privacy, transparency, accountability (Weitzner et al. 2008) and auditing for both Users and regulators. For the Web to be a free and open innovation platform should be engineered as net neutral

(Vafopoulos 2012), not fragmented (e.g. (Yeung et al. 2009)) and uncensored (Kroes 2011) space. Public and private funding to independent institutions should sustain open and effective standards. Furthermore, it is of equal importance that the regulatory policies acquire the right balance of market power and innovation, and favor Web-based development (Vafopoulos 2012; Vafopoulos 2005; Vafopoulos, Gravvanis, and Platis 2006).

The Web ecosystem is a unique opportunity to re-think and re-engineer human life, whereas the prominent role of philosophy is to communicate this challenge not only in the scientific and ethical tradition but also in the research capability.

**Address**

PO BOX 1077 epektasi thermis, Thermi, Thessaloniki, Greece, 57001.

vaf@aegean.gr